\documentclass[preprint,12pt,english]{elsarticle}
\usepackage{subfigure}
\usepackage{amsmath,amsthm,amsfonts,amssymb}
\usepackage{nicefrac}
\usepackage{babel}
\journal{Physics Letters B}

\begin{document}
\begin{frontmatter}
\title{A nonlinear dynamics for the scalar field in Randers spacetime}


\author[ufca]{J. E. G. Silva}


\author[pici]{R. V. Maluf}

\author[pici]{C. A. S. Almeida}

\address[ufca]{Universidade Federal do Cariri (UFCA), Instituto de forma\c{c}\~ao de professores, Rua Oleg\'{a}rio Em\'{i}dio de Ara\'{u}jo, Brejo Santo-Ce ,63.260.000-Brazil}

\address[pici]{Universidade Federal do Cear\'a (UFC), Departamento de F\'isica, Campus do Pici, Fortaleza - CE, C.P. 6030, 60455-760 - Brazil}



\begin{abstract}
We investigate the properties of a real scalar field in the Finslerian Randers spacetime, where 
the local Lorentz violation is driven by a geometrical background vector. We propose a dynamics for the scalar field by a minimal coupling of the scalar field and the Finsler metric. The coupling is intrinsically defined on the Randers spacetime, and it leads to a non-canonical kinetic term for the scalar field. The nonlinear dynamics can be split into a linear and nonlinear regimes, which depend perturbatively on the even and odd powers of the Lorentz-violating parameter, respectively. We analyze the plane-waves solutions and the modified dispersion relations, and it turns out that the spectrum is free of tachyons up to second-order. 
\end{abstract}
\end{frontmatter}

\section{Introduction}

Despite the current lack of a complete theory of quantum gravity, several candidate theories assume that some symmetries present at low energy regimes might no longer be valid at Planck scale. For instance, the string theory \cite{KS}, spacetime noncommutativity \cite{noncommutativegeometry}, Horava-Lifshitz gravity \cite{horava}, loop quantum gravity (LQG) \cite{lqg}, Doubly Special Relativity (DSR) \cite{dsr} and the Very Special Relativity (VSR) \citep{vsr} admit the possibility of absence of Lorentz symmetry for the spacetime. The violation of Lorentz symmetry may be the result of a spontaneous breaking of tensor fields which acquire nonvanishing vacuum expectation values \cite{espontaneous lorentz violation string} or the condensation of a ghost (scalar) field leading to modifications of the dispersion relations \cite{ghostcondensation}. A field theoretical framework to test the Lorentz symmetry is provided by the
Standard Model Extension (SME) \citep{sme}. For a comprehensive review of tests on Lorentz and CPT violation, we indicate the Ref.\cite{rmp}.

The violation of the local Lorentz symmetry can be extended to curved spacetimes by means of the so-called Finsler geometry \citep{kosteleckyfinsler,amelino,gibbons} .
In this anisotropic geometry, the intervals are evaluated by a non-quadratic function, called the Finsler function \citep{chern,shen}. The lack of quadratic restriction provides modified dispersion relation for the fields, a hallmark of
Lorentz violation \citep{Kosteleckycl,girelli,visser}. Applications of Finsler geometry can also be found in optics \citep{finsleroptics} and condensed matter physics \citep{cveticgraphene}.

One of the most important Finsler spacetimes is the Randers spacetime where the anisotropy is driven by a background vector field $a_\mu$ which changes the length of intervals of the spacetime \cite{randers,miron}. The cosmological and astrophysical
effects of the Randers spacetime were analyzed in Refs. \citep{chang,changmdr}. In the context of the SME, the Randers spacetime arises as a kind of bipartite-Finsler space, in the classical point particle Lagrangian for the CPT-Odd fermionic sector \citep{Kosteleckybipartite,euclides}. Other SME-based Finsler spaces can be found in Refs. \citep{smefinsler}.

In this work, we propose a minimal coupling of a real scalar field and the Finsler metric in the Randers spacetime. Unlike the
tangent bundle theories \citep{Asanov,pfeifer}, whose dynamics lies on $TTM$, we propose a position dependent field and Lagrangian. Our dynamics also differs from the osculating method, where the direction-dependent is worked out as a constraint \citep{stavrinos}. For a Finslerian action, we employ an extension of the known as the Shen functional, where the square of the components of the gradient of the field is evaluated with the Finsler metric. As a Finsler volume, we choose the Busemann-Hausdorff volume which provides an anisotropic factor, such as in Bogoslovsky space \citep{bogoslovsky}.

The work is organized as follows. In section \ref{randersspacetime} we review the basic definitions and properties of the Randers spacetime. In section \ref{finsleraction} we propose the Finsler action, obtain the Finslerian equation of motion and
analyze the important regimes. The modified dispersion relation and stability are studied in section \ref{lineardynamicssector} and section \ref{nonlinearregime}, for linear and nonlinear regimes, respectively.
Final remarks, conclusions and perspectives are outlined in section \ref{conclusions}.

\section{Randers Spacetime}
\label{randersspacetime}

In Randers spacetime, given a four-velocity $\dot{x}^{\mu}=dx^{\mu}/dt$, the infinitesimal interval of a worldline is defined by $ds=F_R(x,\dot{x})dt$ where \cite{chang}
\begin{eqnarray}
\label{randersfunction}
F_{R}(x,\dot{x})	&	:=	&	\alpha(x,\dot{x})+\beta(x,\dot{x})\nonumber\\
		&	=	&	\sqrt{-g_{\mu\nu}(x)\dot{x}^{\mu}\dot{x}^{\nu}} + \zeta a_{\mu}(x)\dot{x}^{\mu},
\end{eqnarray}
and $\zeta $ is a real parameter controlling the local Lorentz-violation.

The norm of the background Randers covector $a_{\mu}$ is evaluated with the Lorentzian metric $g$, $a^{2}(x):=g_{\mu\nu}(x)a^{\mu}a^{\nu}$. In this work we adopt the mostly-plus metric convention $(-,+,+,+)$. The perturbative character of the Lorentz violation is encoded in the small value of the linear term, constrained to $0\leq \zeta a <1$.
We assume that $\zeta$ is a constant, bigger than Planck length and with a dimension of length $L$. The Randers background vector is assumed to have mass dimension one, as expected for a background vector field arising from reminiscent quantum gravity effects in four dimensions \cite{vacaru}.

The Randers function can be written as $F_{R}=\sqrt{-g^{R}_{\mu\nu}(x,\dot{x})\dot{x}^\mu \dot{x}^\nu}$, where the
anisotropic timelike Randers metric $g^{R}_{\mu\nu}(x,\dot{x})$ is defined as \cite{girelli,amelino,pfeifer}
\begin{eqnarray}
\label{hessianoffinslerfunction}
g^{R}_{\mu\nu}(x,\dot{x})&:=&-\frac{1}{2}\frac{\partial^{2}F_{R}^{2}(x,y)}{\partial \dot{x}^{\mu}\partial \dot{x}^{\nu}}\nonumber\\
					&=& \frac{F_R}{\alpha}g_{\mu\nu} + \frac{\beta}{\alpha}\tilde{l}_{\mu}\tilde{l}_{\nu} - \zeta(\tilde{l}_{\mu}a_{\nu}+\tilde{l}_{\nu}a_{\mu}) - \zeta^{2}a_{\mu}a_{\nu},
\end{eqnarray}
where $\tilde{l}_{\mu}:=\partial\alpha / \partial\dot{x}^\mu = -g_{\mu\nu}(x)\dot{x}^{\nu}/\alpha$, such that $\tilde{l}_{\mu}\dot{x}^\mu = \alpha$.

We define the action for a free massive particle in the Randers-Finsler spacetime as \cite{randers,girelli}
\begin{equation}
\label{finsleraction}
S^{R}=-m\int_{I}dt F_{R}(x,\dot{x})=-m\int_{I} dt \sqrt{-g_{\mu\nu}^{R}(x,\dot{x})\dot{x}^{\mu}\dot{x}^{\nu}}.
\end{equation}

The free particle action \eqref{finsleraction} is analogous to the Lagrangian of a charged particle in a Lorentzian spacetime with an electromagnetic background vector $a_\mu$.
The canonical momentum is $P_{\mu}^{R}=P_{\mu} - m\zeta a_{\mu},$
where $P_{\mu}:=mg_{\mu\nu}(x)U^{\nu}$ is the Lorentzian conjugate momentum and $U^{\mu}:=\dot{x}^{\mu}/\alpha(\dot{x})$ is
the Lorentzian unitary 4-velocity.
The Finsler metric $g^{R}_{\mu\nu}(x,\dot{x})$ provides a nonlinear duality between the covariant $P^{R}_{\mu}$ and the contravariant $P^{R\mu}$, given by $P^{R}_{\mu}=g_{\mu\nu}^{R}(x,P^{R})P^{R\nu}$ \cite{girelli}.
Thus, the contravariant components of the momentum are given by $P^{R\mu}=m\dot{x}^{\mu}/F(x,\dot{x})$ \cite{girelli}. 
The Finsler function for the covariant vector $P^{R}_\mu $ is $$F_{R}^{*}(x,P^{R}_\mu)    
=\sqrt{-g^{*\mu\nu}(x)P^{R}_\mu P^{R}_\nu} + a^{*\mu}(x)P^{R}_\mu,$$
where $$a^{*\mu}(x)    =    - \zeta\frac{a^{\mu}}{(1\mp \zeta^{2} a^{2})},\ \  g^{*\mu\nu}(x)    = \frac{( 1\mp \zeta^{2} a^{2})g^{\mu\nu} - \zeta^{2} a^{\mu}a^{\nu}}{(1\mp\zeta^{2} a^{2})^{2}}.$$
The signs $-,+$ stand for a timelike and spacelike background vector $a^\mu$ \cite{shen,miron}. The dual Finsler metric $g^{*F}(x,P_{\mu}^{R})$ is defined by \cite{girelli,shen} $$g^{*F \mu\nu}(x,P_{\mu}^{R})=-\frac{1}{2}\frac{\partial^{2}F^{*2}(x,P{\mu}^{R})}{\partial P_{\mu}^{R}\partial P_{\mu}^{R}}
.$$ 

The Finsler metric also provides a deformation of the mass shell, given by \cite{girelli,bogoslovsky,gibbons,vacaru}
\begin{equation}
\label{finslermassshell}
g^{*R \mu\nu}(x,P_{\mu}^{R})P^{R}_{\mu}P^{R}_{\nu}=-m^{2}.
\end{equation}
In Randers spacetime, the modified dispersion relation (MDR) is $g^{\mu\nu}P_{\mu}^{R}P^{R}_{\nu} - 2m\zeta (a\cdot P^{R}) = - (1\mp \zeta^{2} a^{2})m^{2}$,
which is an elliptical hyperboloid of two sheets \cite{euclides}. Considering the momentum 4-vector $P^{R\mu}=(E,\vec{p})$, the dual mass-shell $g^{R}_{\mu\nu}(x,P^{R})P^{R\mu}P^{R\nu}=-m^2$ yields to $(g_{\mu\nu}+\zeta^2 a_\mu a_\nu)P^{R\mu}P^{R\nu}-2\zeta m a_\mu P^{R\mu}=-m^2$. In a flat spacetime $g_{\mu\nu}=\eta_{\mu\nu}$ and for a constant Randers background vector $a^\mu =(-a_0, \vec{a})$, the modified mass-shell is $(1-\zeta^2 a_{0}^2)E^2 + 2\zeta a_0 (m-\zeta (\vec{a}\cdot \vec{p}))E-|\vec{p}|^2 +\zeta (\vec{a}\cdot \vec{p})[2m-\zeta (\vec{a}\cdot \vec{p})]=-m^2$. For a timelike background vector $a^\mu =(-a_0, \vec{0})$, the modified mass-shell lies inside the Lorentz-invariant lightcone, similar to the dispersion relations analysed in the Ref.\cite{ralf}. Further, the asymptote of the deformed mass-shell corresponds to the Randers lightcone. Then, though the particle reaches Lorentzian superluminal velocities, its speed does not exceed the speed of light in the Randers spacetime.

\section{Scalar field dynamics}
\label{scalarfielddynamics}

After analyzing the dynamics of a point particle in the anisotropic Randers spacetime, let us now consider the real scalar field dynamics in this local Lorentz violating spacetime. We assume that the real scalar field is a function of the position $x$ only, i.e., $\Phi=\Phi(x)$.

Consider the action functional defined by a minimal coupling between the scalar field and the dual Finsler metric, namely
\begin{equation}
\label{shenaction2}
S_{\Phi}=-\frac{1}{2}\int d^{4}x \sqrt{-g(x)} \left(1\mp\zeta^{2}a^2(x)\right)^{\frac{5}{2}}\left[g^{F*\mu\nu}(x,d\Phi)\partial_{\mu}\Phi\partial_{\nu}\Phi  + V(\Phi)\right],
\end{equation}
where $V(\Phi)=m^2 \Phi^2$ for the free scalar field and the signs $-,+$ stand for a timelike and spacelike background vector $a$ respectively. The anisotropic volume form $d^{4}x\sqrt{-g(x)}(1\mp\zeta^{2}a^2(x))^{5/2}$, is an extension of the Busemann-Haussdorf volume for the Randers spacetime \cite{busemann,shen}.
It is worthwhile to mention that the Finslerian action in Eq.\eqref{shenaction2} bears some resemblance to the so-called \textit{k-essence} models \cite{kessence}.

The nonquadratic Lagrangian density defined by Eq.\eqref{shenaction2} can be split into two terms
\begin{eqnarray}
\mathcal{L}_{\Phi}=\mathcal{L}^{b}_{\Phi}+\mathcal{L}^{NQ}_{\Phi},
\end{eqnarray}
where $\mathcal{L}^{b}_{\Phi}$ is the bilinear Lagrangian density constructed with the quadratic terms field derivatives $\partial\Phi$ and with the potential term, given by
\begin{eqnarray}
\label{bilinearlagrangian}
\mathcal{L}^{b}_{\Phi}	&	=	&-\frac{1}{2}\Bigg\{\Big[(1\mp\zeta^{2}a^{2})g^{\mu\nu}(x)-2\zeta^{2}a^{\mu}(x)a^{\nu}(x)\Big]\partial_{\mu}\Phi\partial_{\nu}\Phi + \nonumber\\
					& & 	+	 V(\Phi)(1\mp\zeta^2 a^2)^2\Bigg\}\sqrt{-(1\mp\zeta^2 a^2)g},
\end{eqnarray}
and $\mathcal{L}^{NQ}_{\Phi}$ corresponds to the nonquadratic terms in the Lagrangian $\mathcal{L}_{\Phi}$, whose expression is given by
\begin{equation}
\label{nonlinearlagrangian}
\mathcal{L}^{NQ}_{\Phi}=-\zeta (a\cdot \partial\Phi)\sqrt{[-(1\mp\zeta^{2}a^{2})(\partial\Phi)^{2}+\zeta^{2}(a\cdot\partial\Phi)^{2}]}\sqrt{-(1\mp\zeta^2 a^2)g},
\end{equation}
where, $(\partial\Phi)^2:=g^{\mu\nu}(x)\partial_{\mu}\Phi\partial_{\nu}\Phi$ and $a\cdot\partial\Phi:=a^{\mu}\partial_{\mu}\Phi$. Note that for a Lorentzian spacetime, i.e. $\zeta=0$, the nonquadratic Lagrangian density $\mathcal{L}^{NQ}_{\Phi}$
vanishes and the bilinear Lagrangian reduces to the Lorentz-invariant $\mathcal{L}_{\Phi}=-\frac{1}{2}\Big[g^{\mu\nu}(x)\partial_{\mu}\Phi\partial_{\nu}\Phi + 2 V(\Phi)\Big]\sqrt{-g}$.

The bilinear Lagrangian $\mathcal{L}_{\Phi}^{b}$ in \eqref{bilinearlagrangian} can be expanded in powers of $\zeta$ as
\begin{eqnarray}
\mathcal{L}_{\Phi}^{b}	&	=	&	\Bigg\{-\frac{1}{2}\Big[(\partial\Phi)^{2}+V(\Phi)\Big]-\frac{\zeta^{2}}{2}\Bigg[\mp\frac{3}{2}a^2(\partial\Phi)^2 - 2(a\cdot \partial \Phi)^2 \mp \frac{5}{2}a^{2}V(\Phi)\Bigg]\nonumber\\
						&	& +		\mathcal{O}(\zeta^4)\Bigg\}\sqrt{-g}.
\end{eqnarray}
Therefore, the Local Lorentz violating effects arise in second order in $\zeta$ parameter. The LV terms have a similar form of those proposed in the context of the Standard Model Extension (SME) \cite{sme}. Indeed, defining the symmetric and dimensionless tensor $k^{\mu\nu}_{\Phi\Phi}$ as $k^{\mu\nu}_{\Phi\Phi}:=\zeta^2 a^{\mu}a^{\nu}$, the bilinear term can be regarded as a CPT even Lorentz-violating lagrangian of the Higgs sector in the minimal SME \cite{sme}.

Extremizing the Finslerian action \eqref{shenaction2}, the Euler-Lagrange equation {\bf yields the} equation of motion for the field $\Phi$
\begin{equation}
\frac{1}{(1\mp\zeta^{2}a^{2})^{\frac{5}{2}}\sqrt{-g}}\partial_{\mu}\Big[(1\mp\zeta^{2}a^{2})^{\frac{5}{2}}\sqrt{-g}g^{F*\mu\nu}(x,d\Phi)\partial_{\nu}\Phi\Big]=\frac{\partial V(\Phi)}{\partial \Phi}.
\label{shenkleingordonequation}
\end{equation}

The Eq.\eqref{shenkleingordonequation} is a nonlinear Finslerian extension of the Klein-Gordon equation. By defining the nonlinear D'Alembertian operator
\begin{equation}
\Box^{F}\Phi:=\frac{1}{(1\mp\zeta^{2}a^{2})^{\frac{5}{2}}\sqrt{-g}}\partial_{\mu}\Big[(1\mp\zeta^{2}a^{2})^{\frac{5}{2}}\sqrt{-g}g^{F*\mu\nu}(x,d\Phi)\partial_{\nu}\Phi\Big],
\end{equation}
the nonlinear Finslerian Klein-Gordon equation can be rewritten as
\begin{equation}
\Box^{F}\Phi=\frac{\partial V(\Phi)}{\partial \Phi}.
\end{equation}

The nonlinear Finslerian D'Alembertian operator $\Box^{F}\Phi$ is an extension of the so-called Shen Laplacian \cite{shen}, defined on Riemann-Finsler spaces, to Pseudo-Finsler Spacetimes.

The free nonlinear Finslerian Klein-Gordon equation, where $V(\Phi)=m^2 \Phi^2$, can be rewritten as
\begin{eqnarray}
\label{nonlinearfinsleriankleingordonequation}
g^{F*\mu\nu}(x,d\Phi)\partial_{\mu}\partial_{\nu}\Phi + \partial_{\mu}\Bigg[\log \Big[(1\mp\zeta^2 a^2)^{\frac{5}{2}}\sqrt{-g}\Big]g^{F*\mu\nu}(x,d\Phi)\Bigg]\partial_{\nu}\Phi = m^{2}\Phi.
\end{eqnarray}
For a flat spacetime with a constant background field $a$, the anisotropic volume factor $(1\mp\zeta^2 a^2)^{5/2}$ can be absorbed in a change of coordinates and the Klein-Gordon equation \eqref{nonlinearfinsleriankleingordonequation} yields to
\begin{equation}
g^{F*\mu\nu}(d\Phi)\partial_{\mu}\partial_{\nu}\Phi + \partial_{\mu}(g^{F*\mu\nu}(d\Phi))\partial_{\nu}\Phi = m^{2}\Phi.
\end{equation}
Consider the ray approximation, where the wave length $\lambda$ is much smaller then the geometrical characteristic length $L$, i.e., $\lambda<<L$ \cite{padmanabhan}.
A ray \textit{ansatz} for the scalar field has the form \cite{padmanabhan}
\begin{equation}
\label{rayansatz}
\Phi(x):=Re\Big[(a_1 + \epsilon a_2 + \cdots)e^{-i\frac{\psi(x)}{\epsilon}}\Big],
\end{equation}
where $\epsilon:=\lambda/L$ and the phase function $\psi(x)$ is called the eikonal. The differential of the field $\Phi$ \eqref{rayansatz} is given by
\begin{equation}
d\Phi=\frac{1}{\epsilon}A(x)k_{1} + \theta_{1},
\end{equation}
where, $A(x):=Re\Big[-i(a_1 + \epsilon a_2 + \cdots)e^{-i\frac{\psi(x)}{\epsilon}}\Big]$, $k_{1} :=k_{\mu}dx^{\mu}$,
$k_{\mu}:=\partial_{\mu}\psi$ and $\theta_{1}:=\theta_{\mu}dx^{\mu}$, where $\theta_{\mu}:=Re\Big[(\partial_{\mu}(a_1 + \epsilon a_2 + \cdots)e^{-i\frac{\psi(x)}{\epsilon}})\Big]$. At leading order, $g^{F*\mu\nu}(x,d\Phi)=g^{F*\mu\nu}(x,A k)=g^{F*\mu\nu}(x,k)$.
Therefore, the  nonlinear Finslerian Klein-Gordon equation \eqref{shenkleingordonequation} yields the modified dispersion relation
\begin{equation}
\label{leadingordermdr}
g^{F*\mu\nu}(x,k)k_{\mu}k_{\nu}=-m^{2}.
\end{equation}
Then, at leading order, the wave 1-form $k$ modified dispersion relation in \eqref{leadingordermdr} satisfies the point particle
modified mass shell \eqref{finslermassshell}.

\section{Linear dynamics sector}
\label{lineardynamicssector}

The bilinear Lagrangian density $\mathcal{L}_{\Phi}^{b}$ in \eqref{bilinearlagrangian} already exhibits interesting anisotropic modifications in its own. The Euler-Lagrange equations from the bilinear Lagrangian $\mathcal{L}_{\Phi}^{b}$ yields the equation
\begin{eqnarray}
\label{linearkleingordonequation}
&& \frac{\partial_{\mu}\partial^{\mu}\Phi}{(1\mp\zeta^{2}a^{2})}- \frac{2\zeta^{2}}{(1\mp\zeta^{2}a^{2})^{2}}\Big[a^{\mu}a^{\nu}\partial_{\mu}\partial_{\nu}\Phi+ (a\cdot\partial\Phi)\partial_{\mu}a^{\mu}+ a^{\mu}\partial_{\mu}a^{\nu}\partial_{\nu}\Phi \nonumber\\ && +  a(\partial_{\mu}a)\partial^{\mu}\Phi -  \frac{2\zeta^2 a^{\mu}\partial_{\mu}a(a\cdot\partial\Phi)}{(1\mp\zeta^2 a^2)}\Big]
+  \frac{\partial_{\mu}\Big[\log{(\sqrt{-g}(1\mp\zeta^2 a^2)^{\frac{5}{2}})}\Big]}{(1+\zeta^2 a^2)^2}\times \nonumber\\ &&
\Big[(1\mp\zeta^{2}a^{2})\partial^{\mu}\Phi -  2\zeta^{2}(a\cdot\partial\Phi)a^{\mu}\Big]
=  \frac{\partial V(\Phi)}{\partial\Phi}.
\end{eqnarray}

In a flat spacetime $(g^{\mu\nu}=\eta^{\mu\nu})$ for a constant background vector $a^{\mu}$, after a rescaling of the coordinates and field, the linear anisotropic Klein-Gordon equation yields to
\begin{equation}
\label{rescaledlinearfinslerkleingordonequation}
\partial_{\mu}\partial^{\mu}\Phi-\zeta^2 (a^{\mu}a^{\nu}\partial_{\mu}\partial_{\nu}\Phi)=\frac{\partial V(\Phi)}{\partial\Phi}.
\end{equation}

Considering a free massive scalar field, where $V(\Phi)=m^2 \Phi^2$, the Fourier transform \textit{ansatz} for $\Phi$
\begin{equation}
\Phi(x):=\int\frac{d^{4}k}{(2\pi)^{4}}\tilde{\Phi}(k)e^{-i k\cdot x},
\end{equation}
in the Klein-Gordon equation Eq.\eqref{rescaledlinearfinslerkleingordonequation} yields the modified dispersion relation
\begin{equation}
\label{scalarfieldmodifieddispersionrelation}
k^{2}-\zeta^2 (a\cdot k)^2=-m^{2},
\end{equation}
where $k^2 :=\eta^{\mu\nu}k_{\mu}k_{\nu}$, $a^2 := \eta^{\mu\nu}a_{\mu}a_{\nu}$ and $a\cdot k:=\eta^{\mu\nu}a_{\mu}k_{\nu}$.
The modified dispersion relation (MDR) in \eqref{scalarfieldmodifieddispersionrelation} has a {\bf form similar to the modified} particle mass shell. The MDR in the linear regime \eqref{scalarfieldmodifieddispersionrelation} resembles the MDR for the graviton in the so-called Bumblebee model \cite{bumblebee, maluf}.

For $k_{\mu}=(k_0,\vec{k})$, the modified dispersion relation \eqref{scalarfieldmodifieddispersionrelation} takes the form
\begin{equation}
\label{modifieddispersionrelationlinearregime}
(1+\zeta^2 a^2_0)k_0^{2} - 2\zeta^{2}a_{0}k_0(\vec{a}\cdot\vec{k})
- \left[|\vec{k}|^2 + m^{2} - \zeta^2 (\vec{a}\cdot\vec{k})^2\right] =  0.
\end{equation}

From \eqref{modifieddispersionrelationlinearregime}, the relation between the frequency and wave vector is given by
\begin{equation}
\label{frequencyfunction}
k_0(\vec{k})	=\frac{1}{(1+\zeta^2 a^2_0)}\left[\zeta^2 a_{0}(\vec{a}\cdot\vec{k})\pm\sqrt{(|\vec{k}|^2 +  m^2)(1+\zeta^2 a_0^2)-\zeta^2 (\vec{a}\cdot\vec{k})^2}\right].
\end{equation}

One should notice that for an arbitrary configuration of the background vector, the dispersion relations assume the form $k_{0}=E_{\pm}$ with $|E_{+}|\neq|E_{-}|$. This difference between the absolute values of the positive and negative energy states impairs the usual quantum description for particles and antiparticles, leading to problems with respect to the locality of the quantum theory, as discussed in \cite{greenberg,pereira}. Nevertheless, it is interesting to open up the discussion of the spectral consistency of the model for some particular configurations of $a^\mu$.

For $\zeta=0$ we recover the Lorentz-invariant relation $k_{0}=\pm\sqrt{|\vec{k}|^{2}+m^{2}}$. Taking now a timelike background vector $a_{\mu}=(a_0,\vec{0})$, the corresponding dispersion relation is
\begin{equation}
k_0(\vec{k})=\pm\sqrt{\frac{|\vec{k}|^{2}+m^{2}}{1+\zeta^{2}a_{0}^{2}}},
\end{equation}which is like $k_{0}=\pm E$ and, the interpretation of the negative-energy states can be consistently carry out. The group velocity is
\begin{equation}
v_{g}=\sqrt{\frac{1}{1+\zeta^{2}a_{0}^{2}}}\frac{|\vec{k}|}{\sqrt{|\vec{k}|^{2}+m^{2}}},
\end{equation}which is smaller than 1, assuring causality for this mode.

For a spacelike vector, viz., $a_{\mu}=(0,\vec{a})$, the modified dispersion relation can be written as
\begin{equation}
k_0(\vec{k})	=\pm\sqrt{|\vec{k}|^2\left(1-\zeta^{2}|\vec{a}|^{2}\cos^{2}\theta\right) + m^2},
\end{equation}where $\theta$ is the angle between $\vec{a}$ and $\vec{k}$. Again we find a physically acceptable dispersion relation, related with the group velocity
\begin{equation}
v_{g}= \frac{|\vec{k}|(1-\zeta^{2}|\vec{a}|^{2}\cos^{2}\theta)}{\sqrt{m^{2}+|\vec{k}|^{2}(1-\zeta^{2}|\vec{a}|^{2}\cos^{2}\theta)}},
\end{equation}which becomes smaller than 1 for $\zeta^{2}|\vec{a}|^{2}<1$. So, no superluminal signals are created for both configurations.

Finally, we conclude that a lightlike Randers background vector $a^{\mu}$ is inconsistent with the usual quantum interpretation of the negative-energy states as antiparticles. The above results are in accordance with previous studies in the literature \cite{maluf,pereira}.

\section{Nonlinear regime}
\label{nonlinearregime}

Now let us consider the effects of both linear and nonlinear regimes by expanding the equation of motion  \eqref{nonlinearfinsleriankleingordonequation} up to second-order in $\zeta$ in the flat spacetime with constant $a^{\mu}$.
The resulting equation is given by
\begin{eqnarray}
\label{flatkleingordonequationuptosecondorder}
&&\partial_{\mu}\partial^{\mu}\Phi +  \frac{\zeta}{\sqrt{-(\partial\Phi)^2}}\Bigg\{(a\cdot\partial\Phi)\partial_{\mu}\partial^{\mu}\Phi +2a^{\mu}(\partial_{\mu}\partial_{\nu}\Phi)\partial^{\nu}\Phi 
\nonumber\\ && + \frac{(a\cdot\partial\Phi)\partial^{\mu}\Phi\eta^{\rho\sigma}(\partial_{\mu}\partial_{\rho}\Phi)\partial_{\sigma}\Phi}{(-\partial\Phi)^2}\Bigg\} -\zeta^2 (a^{\mu}a^{\nu}\partial_{\mu}\partial_{\nu}\Phi)=m^2 \Phi.
\end{eqnarray}
Considering a single plane wave solution of form $\Phi(x):=Re(e^{-i k\cdot x})
$, the Eq.\eqref{flatkleingordonequationuptosecondorder} has the modified quartic dispersion relation
\begin{equation}
k^4+2\zeta^2 k^2 (a\cdot k)^2 + 2m^2 k^2 - 2\zeta^2 m^2 (a\cdot k)^2 + \zeta^4 (a\cdot k)^4=0.
\end{equation}
Quartic dispersion relation can also be found in the nonminimal SME \cite{smenonminimal}.
For a timelike Randers vector $a_{\mu}=(a_0,\vec{0})$, we obtain
\begin{equation}
\begin{split}
&(1-\zeta^2 a_0^2)^{2}k_{0}^4 - 2\Big[(|\vec{k}|^2 + m^2) - \zeta^2 a_0^2 |\vec{k}|^2\Big]k_0^2 +(|\vec{k}|^2 + m^2)^2=0.
\end{split}
\label{mdrupsecondordertimelike}
\end{equation}
Once again there are two positive $(k_{0;1}^{+},k_{0;2}^{+})$ and two negatives $(k_{0;1}^{-},k_{0;2}^{-})$ roots, where
\begin{equation*}
k_{0;1,2}^{+}= \sqrt{\frac{(1-\zeta^2 a_0^2)|\vec{k}|^2 + m^2 \pm \zeta a_0 m\sqrt{2(1-\zeta^2 a_0^2) |\vec{k}|^2 + (2-\zeta^2 a_0^2)m^2}}{(1-\zeta^2 a_0^2)^{2}}},
\end{equation*}
\begin{equation}
\label{nonlinearfrequency}
k_{0;1,2}^{-}=- \sqrt{\frac{(1-\zeta^2 a_0^2)|\vec{k}|^2 + m^2 \pm \zeta a_0 m\sqrt{2(1-\zeta^2 a_0^2) |\vec{k}|^2 + (2-\zeta^2 a_0^2)m^2}}{(1-\zeta^2 a_0^2)^{2}}}.
\end{equation}

\begin{figure}[h]

\subfigure[Positive frequencies {\bf as functions} of the 3-momentum $\vec{k}$, the perturbed second-order Klein-Gordon equation and a timelike Randers vector. Both the biggest frequency (dashed line) and the smallest frequency (dotted line) behave as the {\bf Lorentz-invariant limit} (thin line).\label{scalarfield_randersFig1}]{
\includegraphics[width=7cm,height=6cm]{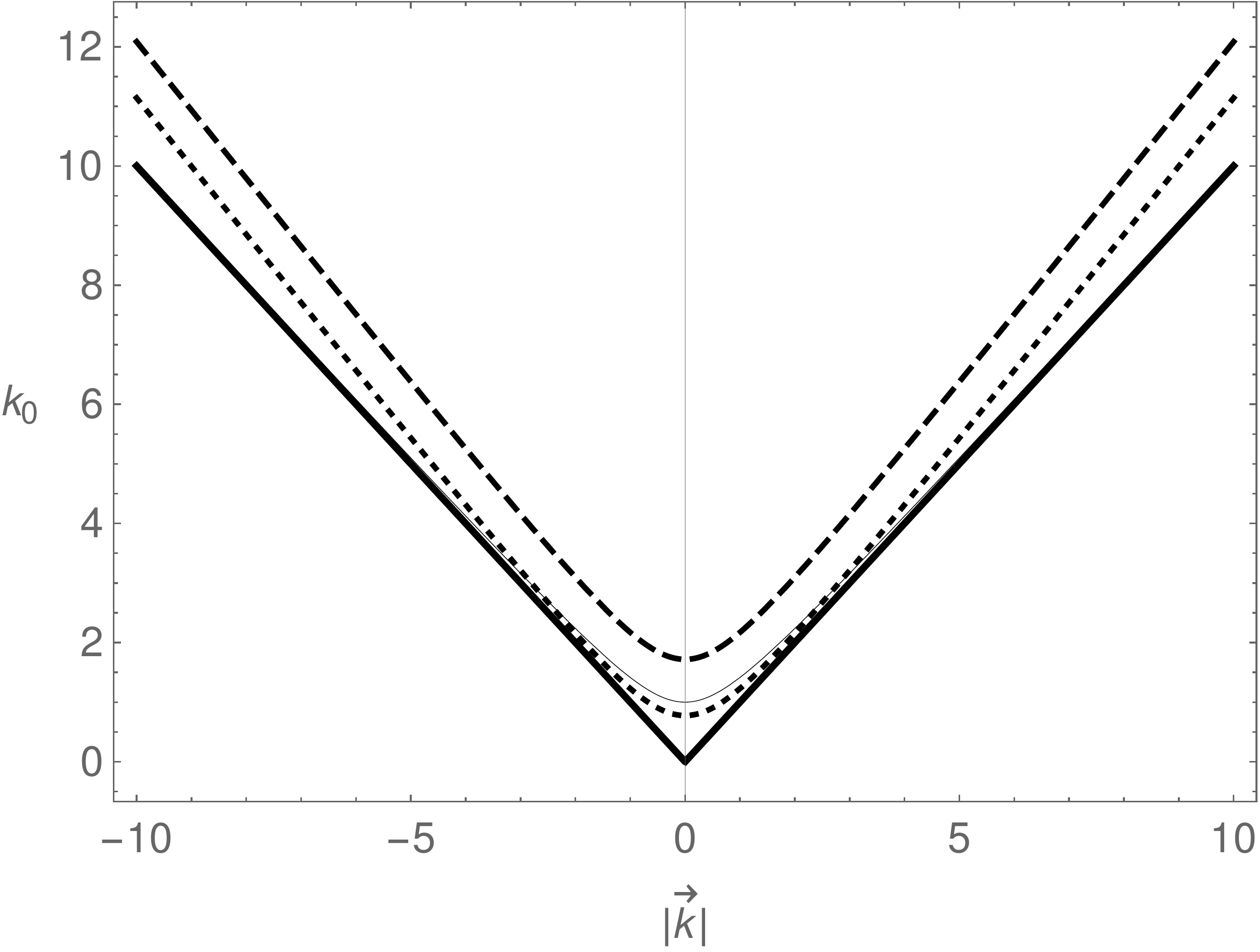}}
\quad
\subfigure[Group velocity for the perturbed second-order Klein-Gordon equation and a timelike Randers vector. The group velocities for the biggest frequency (dashed line) and the smallest frequency (dotted line) {\bf have superluminal behaviour above a $\zeta a_0$-dependent energy scale} Lorentz-invariant one.\label{scalarfield_randersFig2}]{
\includegraphics[width=7cm,height=6cm]{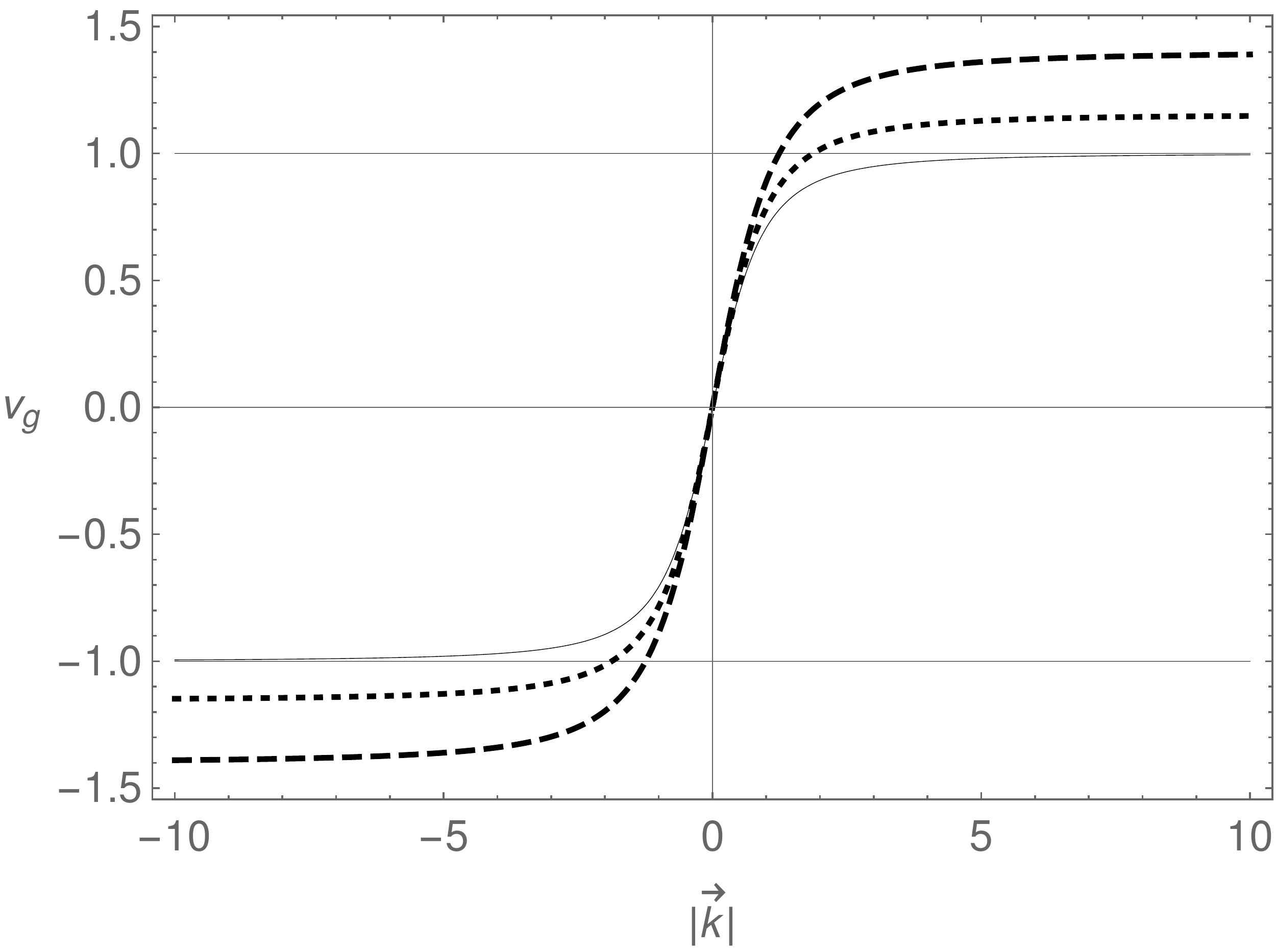}}
\caption{Frequencies and group velocity}
\end{figure}

The behaviour of the frequencies $k_{0;1,2}^{+}$ is shown in the Figure \ref{scalarfield_randersFig1} for $m=1$ and $\zeta a_0 =0.5$. $k_{0;1}^{+}$ (dashed line) and $k_{0;2}^{+}$ (dotted line) are positive what leads to the positivity of the energy.
The absence of complex frequencies provides stability to the states of the scalar field, which means that no exponential decreasing or increasing factor due to the imaginary part appears in the plane-wave, as discussed in details in the Ref.\cite{ralf}. 

The perturbed mass-shells lie inside the Lorentz-invariant lightcone (thick line). The asymptotes represent the deformed Randers lightcone. Further, the difference of the frequencies to the usual frequency (thin line) is proportional to the product $\zeta a_0$ at leading order. 


The group velocity has the form
\begin{equation}
\label{groupvelocitiesupsecondorder}
v_{g;1,2}^{+}		=	\frac{(1-\zeta^2 a_0^2)|\vec{k}|\left[1\pm\zeta a_0 m/\sqrt{(1-\zeta^2 a_0^2)(m^{2}+2|\vec{k}|^2) + m^2}\right]}{\sqrt{(1-\zeta^2 a_0^2 )^{2}|\vec{k}|^2 + m^2 \pm \zeta a_0 m \sqrt{(1-\zeta^2 a_0^2)(m^{2}+2|\vec{k}|^2) +m^2}}},
\end{equation}
\begin{equation}
v_{g;1,2}^{-}=-\frac{(1-\zeta^2 a_0^2)|\vec{k}|\left[1\pm\zeta a_0 m/\sqrt{(1-\zeta^2 a_0^2)(m^{2}+2|\vec{k}|^2) + m^2}\right]}{\sqrt{(1-\zeta^2 a_0^2 )^{2}|\vec{k}|^2 + m^2 \pm \zeta a_0 m \sqrt{(1-\zeta^2 a_0^2)(m^{2}+2|\vec{k}|^2) +m^2}}}.
\end{equation}


\textbf{We plotted the group velocities $v_{g;1,2}^{+}$ in Fig.\ref{scalarfield_randersFig2}. We set $m=1$
and $\zeta a_0=0.5$ (dotted line) and $\zeta a_0=0.7$ (dashed line). The group velocities exhibit superluminal behaviour above an energy scale, in a same way as analysed by Kostelecky and Lehnert in the fermionic sector of the SME \cite{ralf}.}


\section{Conclusions and perspectives}
\label{conclusions}

In this work, we proposed a dynamics for a scalar field intrinsically defined on the local Lorentz-violating Finsler-Randers spacetime.
The action functional proposed is defined through a minimal coupling between the scalar field and the Finsler-Randers metric. We also assume that the anisotropy deforms the volume by means of an extension of the so-called Busemann-Hausdorff Finslerian volume.

The resulting action exhibits a non-canonical kinetic term, as in k-essence models. By expanding the Finslerian action in powers of the background vector,
we obtain Lorentz violating terms similar to those of the Standard Model Extension (SME). The Finslerian equation of motion is an extension of the Klein-Gordon equation by using the so-called Shen D'Alembertian.

The analysis of the \textbf{perturbed} modified dispersion relations (MDR) for the free field revealed that tachyonic modes are absent in the bilinear sector, {\bf regardless of the} causal nature of the background Randers vector. For the nonlinear regime perturbed to second-order,  the fourth-order MDR has a positive energy (stable) spectrum whose group velocities exhibit superluminal effects above an energy scale that depends on $\zeta a_0$. The UV causality issues are similar to those
studies in the fermionic sector of the SME \cite{ralf}. However, since the Randers vector also deforms the lightcone, the particles do not exceed the deformed Randers speed of light.

As perspectives, we point out the analysis of the effects of this nonlinear dynamics in cosmological scenarios. A relevant extension of this work is the analysis of the characteristic surface which could resolve some causality issues at the lightcone. For a complete analysis of the modified dispersion relation, the nonlinear Klein-Gordon equation demands
the use of nonlocal and fractional operators that we leave as a perspective. The quantization of this classical theory and the analysis of some process in order to find upper bounds to the background Randers vector are in progress.

\section{Acknowledgements}
This work was partially supported by the Brazilian agencies Coordena\c{c}\~ao de Aperfei\c{c}oamento de Pessoal de N\'ivel Superior (CAPES) (grant no. 99999.\\006822/2014-02) and  Conselho Nacional de Desenvolvimento Cient\'{\i}fico e Tecnol\'{o}gico (CNPq) (grant numbers 305766/2012-0 and 305678/2015-9). J.E.G. Silva acknowledges the Indiana University Center for Spacetime Symmetries (IUCSS) for the kind hospitality.

\end{document}